\documentclass[12pt,thmsa]{article}
\usepackage{sw20lart}

\input tcilatex
\QQQ{Language}{
American English
}

\begin{document}

\author{I. Radinschi\thanks{%
iradinsc@phys.tuiasi.ro} \and Physics Department, ``Gh. Asachi'' Technical
University, \and Iasi, 6600, Romania}
\title{The KKW Generalized Analysis for a Magnetic Stringy Black Hole}
\date{December 15, 2004}
\maketitle

\begin{abstract}
We apply the Keski-Vakkuri, Kraus and Wilczek (KKW) generalized analysis to
a magnetic stringy black hole solution to compute its temperature and
entropy. The solution that we choose in the Einstein-dilaton-Maxwell theory
is the dual solution known as the magnetic black hole solution.

Our results show that the expressions of the temperature and entropy of this
non-Schwarzschild-type black hole are not the Hawking temperature and the
Bekenstein-Hawking entropy, respectively. In addition, the extremal magnetic
stringy black hole is not frozen because it has a constant non-zero
temperature.


PACS: 04. 20 Dw, 04. 70. Bw,
\end{abstract}

\section{INTRODUCTION}

As we know from the literature, the Keski-Vakkuri, Kraus and Wilczek (KKW)\
analysis [1] has been used to studying the Schwarzschild-type black hole
solution [2] and, after this, to compute the temperature and entropy of
other black hole space-times [3]-[5]. In this analysis, the total
Arnowitt-Desser-Misner mass [6] is fixed but the mass of the Schwarzschild
black hole decreases due to the emitted radiation. To avoid the
singularities at the horizon the Painlev\'{e} [7] coordinate transformation
is used and this enable us the study of the across-horizon physics such as
the black hole radiation. The black hole temperature depends not only on the
characteristics of the black hole but, also, on the energy of the emitted
shell of energy. Furthermore, the black hole entropy is not given by
Bekenstein and Hawking formula for the specific black hole.

For studying the temperature and entropy of a black hole solution which is
not of Schwarzschild-type we can apply successfully the (KKW)\ generalized
analysis. This generalized (KKW) analysis case was studied by Vagenas [8]
and he established the formulas for the temperature and entropy of a black
hole solution described by a metric which satisfies $A(r)\cdot B^{-1}(r)\neq
1$. The Hawking radiation is viewed as a tunneling process which emanates
from the non-Schwarzschild-type black hole solution. Vagenas [8] introduced
a more general coordinate transformation in order to apply the (KKW)\
analysis to non-Schwarzschild-type black holes. He established two
conditions: 1) the regularity at the horizon which ensures that we can study
the across-horizon physics and 2) the stationarity of the non-static metric
which implies that the time direction is a Killing vector, condition that is
very important to generalize the (KKW)\ analysis. The results of this
generalized analysis are the exact expressions of the temperature and
entropy of the non-Schwarzschild-type black holes which are not the Hawking
temperature $T_H$ and the Bekenstein-Hawking entropy $S_{BH}$.

In this paper we apply the Keski-Vakkuri, Kraus and Wilczek (KKW)
generalized analysis to a magnetic stringy black hole solution [9] in order
to evaluate the temperature and entropy of this black hole space-time. The
solution that we study is one of the Einstein-dilaton-Maxwell theory, the
dual solution known as the magnetic black hole solution. The metric is
obtained by multiplying the electric metric in the Einstein frame by a
factor $e^{-2\,\Phi }$.

\section{KKW GENERALIZED ANALYSIS FOR THE MAGNETIC STRINGY BLACK\ HOLE}

String theory may be the best way to attain the holy grail of fundamental
physics, which is to generate all matter and forces of nature from one basic
building block. Through the years, many important studies have been made
related to the string theory.

The low-energy effective theory largely resembles general relativity with
some new ``matter'' fields as the dilaton, axion etc [9]-[10]. A main
property of the low-energy theory is that there are two different frames in
which the features of the space-time may look very different. These two
frames are the Einstein frame and the string frame and they are related to
each other by a conformal transformation ($g_{\mu \nu }^E=e^{-2\,\Phi
}g_{\mu \nu }^S$) which involves the massless dilaton field as the conformal
factor. Even the existence of these two different frames makes the
difference between the two theories, the low energy effective theory and the
Einstein theory. The string frame is actually similar to the Brans-Dicke
frame in the Jordan-Brans-Dicke theory. The string ``sees'' the string
metric. Many of the important symmetries of string theory also rely of the
string frame or the Einstein frame. The $T$ duality [11] transformation
relates metrics in the string frame only, whereas $S$ duality [12] is a
valid symmetry only if the equations are written in the Einstein frame. Kar
[9] gave important results about the stringy black holes and energy
conditions. There were studied several black holes in two and four
dimensions with regard to the Weak Energy Conditions (WEC). It is very
important to study the black holes in string theory and this can be
explained in the following way. Because string theory is expected to provide
us with a finite and clearly defined theory of quantum gravity, the answer
of many questions related with black hole evaporation could be solved in the
context of string theory. To do this, is required to construct black hole
solutions in string theory.

The action for the Einstein-dilaton-Maxwell theory in $3+1$ dimensions is
given by

\begin{equation}
S_{EDM}=\int d^4x\sqrt{-g}e^{-2\,\Phi }[R+4\,g_{ik}\,\nabla ^i\,\Phi
\,\nabla ^k\Phi -\frac 12g^{ij}g^{kl}F_{ik}F_{jl}].  \label{1}
\end{equation}

Varying with respect to the metric, dilaton and Maxwell fields we get the
field equations for the theory given as

\begin{equation}
R_{ik}=-2\,\nabla _i\,\Phi \,\nabla _k\Phi +2\,F_{ij}F_k^{\,\,\,\,j}\,,
\label{2}
\end{equation}

\begin{equation}
\nabla ^k(e^{-2\,\Phi }F_{ik})=0,  \label{3}
\end{equation}

\begin{equation}
4\,\nabla ^2\Phi -4(\nabla \Phi )^2+R-F^2=0.  \label{4}
\end{equation}

The metric (in the string frame) which solve the Einstein-dilaton-Maxwell
field equations to yield the electric black hole is given by

\begin{equation}
ds^2=-A(1+\frac{2\,M\,\sinh ^2\alpha }r)^{-2}\,dt^2=\frac
1A\,dr^2+r^2\,d\theta ^2+r^2\,\sin ^2\theta \,d\varphi ^2.  \label{5}
\end{equation}

where $A=1-\frac{2\,M}r$.

In the string frame the dual solution known as the magnetic black hole is
obtained by multiplying the electric metric in the Einstein frame by a
factor $e^{-2\,\Phi }$. This is in a generalized sense even the $S$-duality
transformation which changes $\Phi \rightarrow -\Phi $ and thereby inverts
the strength of the string coupling. We mention that the magnetic and
electric solutions are the same if looked at from the Einstein frame.
Therefore, the magnetic black hole metric is given by

\begin{equation}
ds^2=-\frac AB\,dt^2+\frac 1{A\,B}\,dr^2+r^2\,d\theta ^2+r^2\,\sin ^2\theta
\,d\varphi ^2,  \label{6}
\end{equation}
where

\begin{equation}
B=1-\frac{Q^2}{M\,r}.  \label{7}
\end{equation}

The metric describes a black hole with an event horizon at $r_{+}=2\,M$.

The (KKW) methodology generalized by Vagenas [8] in the case of the
non-Schwarzschild-type black hole geometries requires that $A(r)\cdot
B^{-1}(r)\neq 1$, condition that is satisfied by the solution given by (6)
and (7). Also, the total Arnowitt-Desser-Misner mass $M_{ADM}$ have to be
well-defined so we get $A(r)\rightarrow 1,$ as $r\rightarrow \infty $ and $%
B(r)\rightarrow 1,$ as $r\rightarrow \infty $. We need to have the
regularity at the event horizon and, also, the stationarity of the
non-static metric which implies that the time direction is a Killing vector
[13]-[14]. The Painlev\'{e} [7] more general coordinate transformation is
given by

\begin{equation}
\sqrt{A(r)}dt=\sqrt{A(r)}d\tau -\sqrt{B^{-1}(r)-1}dr,  \label{8}
\end{equation}

where $\tau $ is the new time coordinate. The metric given by (6) becomes

\begin{equation}
ds^2=-A(r)\,d\tau ^2+2\sqrt{\frac{A(r)}{B(r)}(1-B(r))}\,d\tau
\,dr+dr^2+r^2\,d\theta ^2+r^2\,\sin ^2\theta \,d\varphi ^2.  \label{9}
\end{equation}

The radial null geodesics are

\begin{equation}
\QATOP{.}{r}=\sqrt{\frac{A(r)}{B(r)}}[_{-}^{+}1-\sqrt{(1-B(r))}].  \label{10}
\end{equation}

In the equation above, the upper (lower) sign corresponds to the outgoing
(ingoing) geodesics under the assumption that $\tau $ increases towards
future.

The total Arnowitt-Desser-Misner mass $M_{ADM}$ is fixed and the mass $M$ of
the black hole fluctuates because a shell of energy $\omega $ which
constitutes of massless particle considering only the s-wave part of
emission, is radiated by the black hole. Now, the massless particles travel
on the outgoing geodesics which are due to the varying mass $M$ of the black
hole. The metric is given by

\begin{equation}
\begin{tabular}{c}
$ds^2=-A(r,M-\omega )\,d\tau ^2+2\sqrt{\frac{A(r,M-\omega )}{B(r,M-\omega )}%
(1-B(r,M-\omega ))}\,d\tau \,dr+$ \\ 
\\ 
$+dr^2+r^2\,d\theta ^2+r^2\,\sin ^2\theta \,d\varphi ^2.$%
\end{tabular}
\label{11}
\end{equation}

We get that the outgoing radial null geodesics will have a new formula

\begin{equation}
\QATOP{.}{r}=\sqrt{\frac{A(r,M-\omega )}{B(r,M-\omega )}}[1-\sqrt{%
(1-B(r,M-\omega ))}].  \label{12}
\end{equation}

We make the approximation

\begin{equation}
\sqrt{\frac{A^{^{\prime }}}{B^{^{\prime }}}}(1-\sqrt{1-B^{\prime }})\approx
\frac 12\sqrt{A^{^{\prime }}\,B^{^{\prime }}},  \label{13}
\end{equation}

where $A^{^{\prime }}=A(r,M-\omega ^{^{\prime }})\,$and $B^{^{\prime
}}=B(r,M-\omega ^{^{\prime }})$ and the imaginary part of the action (see
equation (32) in [8]) takes the form

\begin{equation}
\func{Im}I=\func{Im}\int_{r_{+}(M-\omega )}^{r_{+}(M}\int_0^{+\omega }\frac{%
d\omega ^{^{\prime }}}{\sqrt{\frac{A^{^{\prime }}}{B^{^{\prime }}}}(1-\sqrt{%
1-B^{^{\prime }}})}dr.  \label{14}
\end{equation}

Now, we calculate the temperature of the black hole (see equation (35) in
[8]) and we get

\begin{equation}
T_{bh}(M,\omega )=\frac \omega {4\,\pi \,M^2[1-(1-\frac \omega M)^2]}.
\label{15}
\end{equation}

The expression of the entropy is given by

\begin{equation}
S_{bh}=S_{BH}-4\,\pi \,M^2[1-(1-\frac \omega M)^2].  \label{16}
\end{equation}

If we evaluate the $T_{bh}$ temperature to first order in $\omega $ we get
the Hawking temperature $T_H$ of the magnetic stringy black hole which is

\begin{equation}
T_H=\frac 1{8\,\pi \,M}.  \label{17}
\end{equation}

The entropy of the black hole is not the Bekenstein-Hawking entropy $S_{BH}$

\begin{equation}
S_{BH}=\pi \,r_{+}^2=4\,\pi \,M^2.  \label{18}
\end{equation}

Furthermore, the black hole entropy $S_{bh}$ to zeroth order in $\omega $
yields the Bekenstein-Hawking entropy $S_{BH}$. We make the calculations in
a new framework, where the extremal magnetic stringy black hole which is
created when we have

\begin{equation}
Q^2=2(M-\omega )^2,  \label{19}
\end{equation}

so, the extremality condition $r_{+}=r_{-}$ is not the same and the
temperature of the extremal black hole does not vanish and becomes

\begin{equation}
T_{bh}(M,Q)=\frac 1{4\,\pi \,M(1+\frac Q{\sqrt{2}\,M})}.  \label{20}
\end{equation}

Also, the following condition has to be respect

\begin{equation}
Q<\sqrt{2}\,M,  \label{21}
\end{equation}

because the emitted shell of energy $\omega $ has to take only positive
values. A naked singularity will not appear from the collapse of the
magnetic stringy black hole.

\section{DISCUSSION}

We used the (KKW) generalized analysis given introduced in [8] in the case
of a non-Schwarzschild-type black hole solution which is a magnetic stringy
black hole solution in the Einstein-dilaton-Maxwell theory, in order to
evaluate the temperature and entropy of this black hole space-time. The
temperature and entropy of the black hole are different from the Hawking
temperature $T_H$ and the Bekenstein-Hawking entropy $S_{BH}$. The black
hole temperature depends on the emitted massless particle's energy. The
temperature and entropy of the black hole in the first and, respectively,
zeroth orders of $\omega $ correspond to the Hawking temperature $T_H$ and
the Bekenstein-Hawking entropy $S_{BH}$. These results sustain the validity
of the generalized KKW analysis introduced by Vagenas [8].

We established that the extremal magnetic stringy black hole is not frozen.
Also, it has a non-zero background temperature since the extremality
condition was changed.

\end{document}